\documentclass{ifacconf}

\usepackage{graphicx}      
\usepackage{natbib}        
\usepackage{lipsum}
\usepackage{siunitx}
\usepackage{amsmath}
\usepackage{amssymb}
\usepackage{bm}
\usepackage{tikz}
\usetikzlibrary{positioning}
\usetikzlibrary{shapes}
\usepackage{tabularx}
\usepackage{booktabs}
\usepackage{subcaption}
\usepackage{pgfplots}
\newcolumntype{s}{>{\hsize=.5\hsize}X}

\newcommand{\eg}{e.\,g.}
\usepackage{eurosym}
\DeclareSIUnit{\sieuro}{\mbox{\euro}}

\newcommand{\intd}{\mathrm{d}}
\newcommand{\inp}{\mathrm{in}}
\newcommand{\out}{\mathrm{out}}
\newcommand{\soil}{\mathrm{s}}
\newcommand{\air}{\mathrm{a}}
\newcommand{\pipe}{\mathrm{pi}}
\newcommand{\cvcv}{\mathrm{cv}}
\newcommand{\water}{\mathrm{w}}
\newcommand{\hydraulic}{\mathrm{h}}
\newcommand{\temp}{\mathrm{t}}
\newcommand{\control}{\mathrm{c}}
\newcommand{\elec}{\mathrm{el}}

\newcommand{\differential}{\mathrm{de}}
\newcommand{\nominal}{\mathrm{nom}}

\newcommand{\ve}[1]{\bm{#1}}
\newcommand{\dve}[1]{\dot{\bm{#1}}}
\newcommand{\hve}[1]{\hat{\bm{#1}}}
\newcommand{\transve}[1]{\bm{#1}^\top}
\newcommand{\ma}[1]{\bm{#1}}

\usetikzlibrary{shapes.geometric, arrows}
\tikzstyle{startstop} = [rectangle, rounded corners, 
minimum width=1cm, 
minimum height=1cm,
text centered, 
draw=black, 
fill=red!30]

\tikzstyle{io} = [trapezium, 
trapezium stretches=true, 
trapezium left angle=70, 
trapezium right angle=110, 
minimum width=3cm, 
minimum height=1cm, text centered, 
draw=black, fill=blue!30]

\tikzstyle{process} = [rectangle, 
minimum width=2cm, 
minimum height=0.5cm, 
text centered, 
text width=2.8cm, 
draw=black, 
fill=orange!30]

\tikzstyle{decision} = [diamond, 
minimum width=1.2cm, 
minimum height=0.25cm,
align=center,
aspect=2,
text width=1.2cm, 
draw=black, 
fill=green!30]
\tikzstyle{arrow} = [thick,->,>=stealth]

\definecolor{airtemp}{HTML}{508AAD}
\definecolor{soiltemp}{HTML}{764134}

\usetikzlibrary{plotmarks}
\usetikzlibrary{arrows.meta}
\usepgfplotslibrary{patchplots}
\usepackage{grffile}
\usepgfplotslibrary{groupplots}
\begin{document}
\begin{frontmatter}

\title{Model Predictive Control of Thermo-Hydraulic Systems Using Primal Decomposition
\thanksref{footnoteinfo}} 

\thanks[footnoteinfo]{This research is supported by the German federal ministry of economic affairs and climate action (BMWK) under the agreement no. 03EWR007O2.}
\author[First]{Jonathan Vieth}
\author[Second,Third]{Annika Eichler}
\author[First]{Arne Speerforck}

\address[First]{Institute of Engineering Thermodynamics, Hamburg University of Technology, Denickestraße 17, 21073 Hamburg, Germany (e-mail: jonathan.vieth@tuhh.de, arne.speerforck@tuhh.de).}
\address[Second]{Institute of Control Systems, Hamburg University of Technology, Hamburg, Germany (e-mail: annika.eichler@tuhh.de)}
\address[Third]{Deutsches Elektronen-Synchrotron DESY, Hamburg, Germany (e-mail: annika.eichler@desy.de)}

\begin{abstract}                
Decarbonizing the global energy supply requires more efficient heating and cooling systems. Model predictive control enhances the operation of cooling and heating systems but depends on accurate system models, often based on control volumes. We present an automated framework including time discretization to generate model predictive controllers for such models. To ensure scalability, a primal decomposition exploiting the model structure is applied. The approach is validated on an underground heating system with varying numbers of states, demonstrating the primal decomposition's advantage regarding scalability.
\end{abstract}


\end{frontmatter}

\section{INTRODUCTION}
The \cite{european_commission_european_2019} mandates that building heat supply in the EU must be decarbonized by 2050. A study by \cite{fraunhofer_institute_for_solar_energy_systems_wege_2021} indicates that the share of buildings heated by district heating networks (DHNs) in Germany must increase to meet climate targets, aligning with \cite{lund_role_2010}. To meet the climate targets, DHNs must be decarbonized.
Decarbonizing DHNs requires improving efficiency through strategies like intelligent operation, see \cite{lund_4th_2014}.
Additionally, cooling demand is the fastest-growing energy use globally; efficiency measures could reduce related electricity costs by about \SI{38}{\percent} compared to baseline scenarios, according to \cite{iea_future_2018}. In summary, the need for efficiency measures in heating and cooling systems will grow significantly.

Efficiency can be improved through optimized planning and operations. \cite{maurer_toward_2023} study optimal DHN operation using model predictive control (MPC), while \cite{wack_economic_2023} address DHN planning. \cite{deng_minlp_2017} apply optimization to district heating and cooling networks, which combine heat and cooling supply using soil as natural storage. \cite{taheri_model_2022} review MPC for heating, ventilation, and air conditioning systems, highlighting its growing attention recently.


All these systems share two features: they can be modeled using control volumes (CVs), and fluid typically circulates in pipes to transport energy. \cite{westphal_enabling_2025} present a simulation library for DHNs. This library is used by \cite{vieth_district_2025} for optimal DHN planning. \cite{maurer_comparison_2021} compare pipe models for DHN MPC, concluding that CV-based models are preferable due to simplicity and a fixed number of variables, unlike the Node Method by \cite{benonysson_dynamic_1991}. However, CV models result in systems described by time-continuous differential-algebraic equations (DAEs).


Automated approaches for MPC generation for time-continuous DAEs are presented in \cite{fabien_dsoa_2010}, \cite{houska_acado_2011}, and \cite{chen_matmpc_2019}. In \cite{fabien_dsoa_2010}, the optimal control problem is solved using a direct method. According to \cite{olanrewaju_implications_2017}, in direct methods first the time-continuous DAEs are time-discretized, and then an optimization problem is solved based on the discretized system model. In contrast, the approaches presented by \cite{houska_acado_2011} and \cite{chen_matmpc_2019} employ multiple shooting techniques to solve the optimal control problem based on the continuous system model.

This work presents a direct method for thermo-hydraulic systems modeled by CVs. Although time discretization of nonlinear systems is challenging, performing it beforehand can improve computational efficiency. Focusing on thermo-hydraulic systems also enables exploiting their specific model structure through primal decomposition.

\subsubsection{Contribution}
In this work, we present an automated workflow for MPC generation consisting of a general dynamic system model for thermo-hydraulic systems (Section~\ref{sec:ContModel}), a procedure for time discretization of nonlinear dynamic systems based on Backward-Differentiation (Section~\ref{sec:DiscreteModel}), and a fast and scalable MPC approach for thermo-hydraulic systems using primal decomposition (Section~\ref{sec:MPC}).

\subsubsection{Notation}
 The identity matrix with $n$ rows and columns is represented by $\ma{I}_n$ and $\ma{1}_{n\times m}$ is a matrix with $n$ rows and $m$ columns where all elements are equal to one. Vectors are denoted by lowercase bold letters, and matrices by uppercase bold letters.
\section{DYNAMICS OF SYSTEMS MODELED BY CONTROL VOLUMES} \label{sec:ContModel}

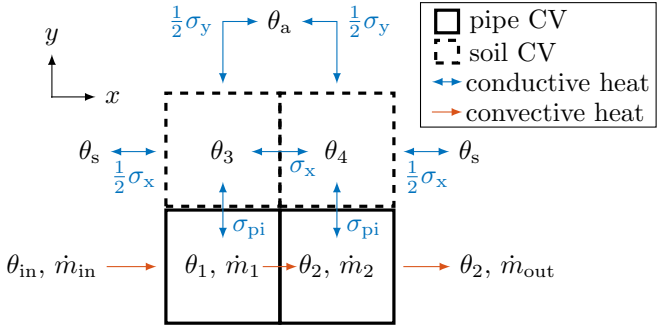
\begin{figure}[t!]
  \centering
  \definecolor{mycolor1}{rgb}{0.00000,0.44700,0.74100}%
\definecolor{mycolor2}{rgb}{0.85000,0.32500,0.09800}%
\definecolor{mycolor3}{rgb}{0.92900,0.69400,0.12500}%

\begin{tikzpicture}
    \draw[black, very thick] (0,0) rectangle (1.5,1.5);
    \draw[black, very thick] (1.5,0) rectangle (3,1.5);
    \draw[black, very thick, dashed] (0,1.55) rectangle (1.5,3.05);
    \draw[black, very thick, dashed] (1.5,1.55) rectangle (3,3.05);
    \node (CV1_left) at (0,0.75) {};
    \node (CV2_right) at (3,0.75) {};
    \node (CV1_right) at (1.15,0.75) {};
    \node (CV2_left) at (1.85,0.75) {};
    \node (CV3_left) at (0,2.275) {};
    \node (CV3_right) at (1,2.275) {};
    \node (CV4_right) at (3,2.275) {};
    \node (CV4_left) at (2,2.275) {};
    \node (CV1_top) at (0.75,1) {};
    \node (CV2_top) at (2.25,1) {};
    \node (CV3_top) at (0.75,3.05) {};
    \node (CV4_top) at (2.25,3.05) {};
    \node (CV3_bot) at (0.75,2) {};
    \node (CV4_bot) at (2.25,2) {};
    \node (cv1) at (0.75,0.75) {$\theta_1$, $\dot{m}_1$};
    \node (cv2) at (2.25,0.75) {$\theta_2$, $\dot{m}_2$};
    \node (cv3) at (0.75,2.275) {$\theta_3$};
    \node (cv4) at (2.25,2.275) {$\theta_4$};
    \node (m_dot_in) at (-1.5,0.75) {$\theta_{\inp}$, $\dot{m}_{\inp}$};
    \node (air) at (1.5,4) {$\theta_{\air}$};
    \node (soil1) at (-1,2.275) {$\theta_{\soil}$};
    \node (soil2) at (4,2.275) {$\theta_{\soil}$};
    \node (m_dot_out) at (4.5,0.75) {$\theta_2$, $\dot{m}_{\out}$};
    \draw[->,>=latex,mycolor2] (m_dot_in.east)  -- (CV1_left);
    \draw[->,>=latex,mycolor2] (CV2_right)  -- (m_dot_out.west);
    \draw[->,>=latex,mycolor2] (CV1_right)  -- (CV2_left);
    \draw[<->,>=latex,mycolor1] (air.east)  -| node[anchor=west] {$\frac{1}{2}\sigma_{\mathrm{y}}$} (CV4_top.north);
    \draw[<->,>=latex,mycolor1] (air.west)  -| node[anchor=east] {$\frac{1}{2}\sigma_{\mathrm{y}}$} (CV3_top.north);
    \draw[<->,>=latex,mycolor1] (CV3_right)  -- node[anchor=north west] {$\sigma_{\mathrm{x}}$} (CV4_left);
    \draw[<->,>=latex,mycolor1] (CV1_top)  -- node[anchor=north west] {$\sigma_{\pipe}$} (CV3_bot);
    \draw[<->,>=latex,mycolor1] (soil1)  -- node[anchor=north] {$\frac{1}{2}\sigma_{\mathrm{x}}$} (CV3_left);
    \draw[<->,>=latex,mycolor1] (soil2)  -- node[anchor=north] {$\frac{1}{2}\sigma_{\mathrm{x}}$} (CV4_right);
    \draw[<->,>=latex,mycolor1] (CV2_top)  -- node[anchor=north west] {$\sigma_{\pipe}$} (CV4_bot);

    \draw[black] (3.35,2.6) rectangle (6.45,4.25);
    \node[rectangle, draw=black,very thick, minimum size = 3mm] (pipeCV) at (3.7,4) {};
    \node (pipeCVtext) [right=of pipeCV,xshift=-1cm] {pipe CV};
    \node[rectangle, draw=black,very thick, minimum size = 3mm,dashed] (soilCV) at (3.7,3.6) {};
    \node (soilCVtext) [right=of soilCV,xshift=-1cm] {soil CV};
    \draw[<->,>=latex,mycolor1] (3.5,3.2) -- (3.9,3.2);    
    \node (cond) at (3.7,3.2) {};
    \node (condText) [right=of cond,xshift=-1cm] {conductive heat};
    \draw[->,>=latex,mycolor2] (3.5,2.8) -- (3.9,2.8);    
    \node (conv) at (3.7,2.8) {};
    \node (convText) [right=of conv,xshift=-1cm] {convective heat};

    \node (axisOrigin) at (-1.5,3) {};
    \node (axisX) at (-.7,3) {$x$};
    \draw[->,>=latex] (-1.512,3) -- (axisX);
    \node (axisZ) at (-1.5,3.8) {$y$};
    \draw[->,>=latex] (-1.5,2.989) -- (axisZ);
\end{tikzpicture}
  \caption{Simple setup of CVs}
  \label{fig:CV}
\end{figure}
\begin{subequations} \label{eq:systemDynSimple}
Consider the system shown in Figure~\ref{fig:CV}, which could represent any type of heating or cooling system. The differential equations (DEs)
\begin{align}
    \frac{\intd \ve{\theta}}{\intd t} = \ma{A} \ve{\theta} + \ma{D} \ve{d} + \ma{X}_{\mathrm{\theta}} \left[ \dve{m} \circ \left( \ma{Z}_{\mathrm{\theta}} \ve{\theta} \right) \right] + \ma{X}_{\mathrm{z}} \left[\ma{Y}_{\mathrm{z}} \dve{m} \theta_{\inp} \right] \label{eq:systemDynSimple_sys}
\end{align}
describe the temperatures $\ve{\theta}^{\top}=\begin{bmatrix}
    \theta_1 & \theta_2 & \theta_3 & \theta_4
\end{bmatrix}$ and the mass flows $\dve{m}^{\top}=\begin{bmatrix}
    \dot{m}_1 & \dot{m}_2 \end{bmatrix}$ in the CVs. The vector $\ve{d}^{\top}=\begin{bmatrix}
    \theta_{\soil} & \theta_{\air}
\end{bmatrix}$ comprises the non-controllable inputs: the soil temperature $\theta_{\soil}$ and the air temperature $\theta_{\air}$. Here, the input temperature $\theta_{\inp}$ is assumed to be an algebraic state (later denoted by $z$), as in many heating or cooling systems it would be linked with $\theta_2$ and an input variable via an algebraic equation.
The constant matrices in \eqref{eq:systemDynSimple_sys} are
\begin{align*}
    \ma{A} = \begin{bmatrix}
        - \frac{\sigma_{\pipe}}{c_{\water} \rho_{\water}V} & 0 & \frac{\sigma_{\pipe}}{c_{\water} \rho_{\water}V} & 0 \\
        0 & - \frac{\sigma_{\pipe}}{c_{\water} \rho_{\water}V} & 0 & \frac{\sigma_{\pipe}}{c_{\water} \rho_{\water}V} \\
        \frac{\sigma_{\pipe}}{c_{\soil} \rho_{\soil}V} & 0 & -\frac{\sigma_{\pipe}+\frac{3}{2}\sigma_{\mathrm{x}}+\frac{1}{2}\sigma_{\mathrm{y}}}{c_{\soil} \rho_{\soil}V} & \frac{\sigma_{\mathrm{x}}}{c_{\soil} \rho_{\soil}V} \\
        0 & \frac{\sigma_{\pipe}}{c_{\soil} \rho_{\soil}V} & \frac{\sigma_{\mathrm{x}}}{c_{\soil} \rho_{\soil}V} &  -\frac{\sigma_{\pipe}+\frac{3}{2}\sigma_{\mathrm{x}}+\frac{1}{2}\sigma_{\mathrm{y}}}{c_{\soil} \rho_{\soil}V} \\
    \end{bmatrix},
\end{align*}
\begin{align*}
    \ma{D} = \frac{1}{2}\begin{bmatrix}
        0 & 0 \\
        0 & 0 \\
        \frac{\sigma_{\mathrm{x}}}{c_{\soil} \rho_{\soil}V} & \frac{\sigma_{\mathrm{y}}}{c_{\soil} \rho_{\soil}V} \\
        \frac{\sigma_{\mathrm{x}}}{c_{\soil} \rho_{\soil}V} & \frac{\sigma_{\mathrm{y}}}{c_{\soil} \rho_{\soil}V} \\
    \end{bmatrix}, 
    \ma{X}_{\mathrm{\theta}} = \begin{bmatrix} 
        \frac{1}{\rho_{\water} V} & 0 \\
        0 & \frac{1}{\rho_{\water} V} \\
        0 & 0 \\
        0 & 0 \\
    \end{bmatrix}, 
    \ma{X}_{\mathrm{z}} = \begin{bmatrix} 
        \frac{1}{\rho_{\water} V} \\
        0  \\
        0  \\
        0  \\
    \end{bmatrix},
\end{align*}
\begin{align*}
    \ma{Z}_{\mathrm{\theta}} = \begin{bmatrix}
        -1 & 0 & 0 & 0\\
        1 & -1 & 0 & 0 \\        
    \end{bmatrix}, 
    \ma{Y}_{\mathrm{z}} = \begin{bmatrix}
        1 & 0 \\
    \end{bmatrix}.
\end{align*}
Here, $V$ is the volume of each CV, $\sigma_{\mathrm{x}}$, $\sigma_{\mathrm{y}}$, and $\sigma_{\mathrm{pi}}$ are thermal conductivities, and $c_i$ and $\rho_i$ are the heat capacity and density of the soil ($\soil$) and the water ($\water$) inside the pipe, respectively, all considered as constant, as commonly done for MPCs (\eg\ \cite{maurer_toward_2023}) and simulation models (\eg\ \cite{westphal_enabling_2025}).  In \eqref{eq:systemDynSimple_sys}, we assume that changes in enthalpy due to changes in pressure are negligible, a condition that is generally met for the systems under consideration.  Additionally, algebraic equations (AEs)
\begin{align}
    \dot{m}_{\inp} = \dot{m}_1 = \dot{m}_2 = \dot{m}_{\out} \label{eq:mass_flow_small}
\end{align}
\end{subequations}
connect the mass flow variables. Note that inserting \eqref{eq:mass_flow_small} into \eqref{eq:systemDynSimple_sys} would simplify \eqref{eq:systemDynSimple_sys}. However, the structure of \eqref{eq:systemDynSimple_sys} leads us to the DAE system
\begin{subequations} \label{eq:timeContinuous}
    \begin{align}
        \frac{\intd \ve{\theta}}{\intd t} &= \ma{A} \ve{\theta} + \ma{B} \ve{u} + \ma{C} \ve{z} + \ma{D} \ve{d} + \sum_{\ve{v} \in \mathcal{V} }\ma{X}_{v} \left[ \left( \ma{Y}_{v} \dve{m} \right) \circ \left( \ma{Z}_{v} \ve{v} \right)  \right], \label{eq:timeContinuous_DE} \\
        \ve{0} &= \ve{f}_{\hydraulic}(\dve{m},\ve{z}_{\hydraulic},\ve{u}_{\hydraulic},\ve{d}), \quad
        \ve{0} = \ve{f}_{\temp}(\ve{\theta},\dve{m},\ve{z},\ve{u},\ve{d}), \label{eq:timeContinuousAE} \\
        \ve{0} &\geq \ve{g}_{\hydraulic}(\dve{m},\ve{z}_{\hydraulic},\ve{u}_{\hydraulic},\ve{d}), \quad
        \ve{0} \geq \ve{g}_{\temp}(\ve{\theta},\dve{m},\ve{z},\ve{u},\ve{d}), \label{eq:timeContinuous_inequality}
    \end{align}
\end{subequations}
which represents an arbitrary system of interconnected CVs with the controllable inputs $\ve{u}^{\top}=\begin{bmatrix}
    \transve{u}_{\hydraulic} & \transve{u}_{\temp}
\end{bmatrix}$ and the algebraic variables $\ve{z}^{\top}=\begin{bmatrix}
    \transve{z}_{\hydraulic} & \transve{z}_{\temp}
\end{bmatrix}$. The subscript h denotes variables related to the hydraulics 
and t to the thermal-model 
. All matrices are constant, $\ve{f}_{\hydraulic}$ is a function representing the hydraulic AEs, $\ve{f}_{\temp}$ all other AEs, $\ve{g}_{\hydraulic}$ the hydraulic inequality constraints, $\ve{g}_{\temp}$ all other inequality constraints, and $\mathcal{V} = \{ \ve{\theta},\ve{z},\ve{u},\ve{d} \}$. 
The reason for extracting the hydraulic constraints from the other constraints is solely for notational purposes and does not lead to any modeling restrictions, since all variables can occur in $\ve{f}_{\temp}$ and $\ve{g}_{\temp}$. This approach is motivated by the primal decomposition scheme, as discussed in Section~\ref{sub:primalMPC}.
\section{TIME DISCRETIZATION OF SYSTEMS MODELED BY CONTROL VOLUMES}  \label{sec:DiscreteModel}

This section aims to derive a time-discrete representation of \eqref{eq:timeContinuous_DE}. 
Common nonlinear system discretization methods include Taylor approximation (\cite{kazantzis_time-discretization_1999}), Runge-Kutta integration (\cite{herty_implicit-explicit_2013}; \cite{frey_gaussnewton_2024}), and backward differentiation (\cite{vaclavek_pmsm_2013}). As shown in \cite{vaclavek_pmsm_2013}, Taylor and Runge-Kutta methods can introduce complex nonlinear terms, potentially altering the original structure of \eqref{eq:timeContinuous_DE} and making the system non-bilinear as the discretization order increases. This increases MPC optimization complexity and eliminates the possibility of linear problems via primal decomposition if nonlinear equations that are not bilinear result from the time-discretization approach.

For this reason, second-order backward differentiation is used. According to \cite{alikhani_adaptive_2016}, the time-discrete system dynamics are given by 
\begin{align}
    \begin{split}
        \ve{0}=&\ve{f}_{\differential,k}\left(\ve{\theta}_k,\dve{m}_k,\ve{z}_k,\ve{u}_k,\ve{d}_k\right) \\
        = &\left[\ma{I}_{|\ve{\theta}_k|} - \frac{2}{3} \Delta t \ma{A} \right] \ve{\theta}_{k} - \frac{4}{3} \ve{\theta}_{k-1} + \frac{1}{3} \ve{\theta}_{k-2} \\ - &\frac{2}{3} \Delta t \bigg[\ma{B} \ve{u}_{k} + \ma{C} \ve{z}_{k} + \ma{D} \ve{d}_{k} \\ + &\sum_{\ve{v} \in \mathcal{V} }\ma{X}_{\ve{v}} \left[ \left( \ma{Y}_{\ve{v}} \dve{m}_{k} \right) \circ \left( \ma{Z}_{\ve{v}} \ve{v}_{k} \right)\right]  \bigg] \label{eq:timeDiscrete_DE}
    \end{split}
\end{align}
with the constant timestep $\Delta t$.
Note that the implementation of higher-order backward differentiation is similar.
\section{PRIMAL-DECOMPOSITION OF SYSTEMs MODELED BY CONTROL VOLUMES}  \label{sec:MPC}

\subsection{General MPC optimization problem}
The goal of the MPC is to minimize an objective function
\begin{subequations} \label{eq:mpc:objective_all}
\begin{align}
\begin{split}
    J_k\left(\hve{\theta}_k, \hat{\dve{m}}_k, \hve{z}_k, \hve{u}_k\right) =  &J_{\hydraulic,k}\left(\hat{\dve{m}}_k, \hve{z}_{\hydraulic,k}, \hve{u}_{\hydraulic,k}\right) \\ + &J_{\temp,k}\left(\hve{\theta}_k, \hat{\dve{m}}_k, \hve{z}_k, \hve{u}_k\right)\label{eq:mpc:objective},
\end{split}
\end{align}
with the notation convention
\begin{align*}
    \hve{a}_k = \begin{bmatrix} \transve{a}_k & \transve{a}_{k+1} & \dots & \transve{a}_{k + n_{\control}-1} \end{bmatrix}^\top,
\end{align*}
and with
\begin{align}
    J_{\hydraulic,k}\left(\hat{\dve{m}}_k, \hve{z}_{\hydraulic,k}, \hve{u}_{\hydraulic,k}\right) &= \sum_{\kappa\in\mathcal{K}} j_{\hydraulic,\kappa}\left(\dve{m}_{\kappa}, \ve{z}_{\hydraulic,\kappa}, \ve{u}_{\hydraulic,\kappa}\right), \\
    J_{\temp,k}\left(\hve{\theta}_k, \hat{\dve{m}}_k, \hve{z}_k, \hve{u}_k\right) &= \sum_{\kappa\in\mathcal{K}} j_{\temp,\kappa}\left(\ve{\theta}_{\kappa}, \dve{m}_{\kappa}, \ve{z}_{\kappa}, \ve{u}_{\kappa}\right)
\end{align}
\end{subequations}
over all time-steps $\mathcal{K}=\{k,k+n_{\control}-1\}$ within the control-horizon $n_{\control}$ with respect to the hydraulic constraints
\begin{subequations} \label{eq:mpc:constraintsHydraulic}
\begin{align}
    \ve{f}_{\hydraulic,\kappa}(\dve{m}_{\kappa},\ve{z}_{\hydraulic,\kappa},\ve{u}_{\hydraulic,\kappa},\ve{d}_{\kappa}) &= 0,\quad \forall \kappa \in \mathcal{K}, \label{eq:mpc:alg_hyd} \\
    \ve{g}_{\hydraulic,\kappa}(\dve{m}_{\kappa},\ve{z}_{\hydraulic,\kappa},\ve{u}_{\hydraulic,\kappa},\ve{d}_{\kappa}) &\leq 0,\quad \forall \kappa \in \mathcal{K}, \label{eq:mpc:ineq_hyd}
\end{align}
\end{subequations}
and the non-hydraulic constraints
\begin{subequations} \label{eq:mpc:constraintsTemp}
\begin{align}
    \ve{f}_{\differential,\kappa}\left(\ve{\theta}_{\kappa},\dve{m}_{\kappa},\ve{z}_{\kappa},\ve{u}_{\kappa},\ve{d}_{\kappa}\right) &= 0,\quad \forall \kappa \in \mathcal{K}, \label{eq:mpc:diff} \\
    \ve{f}_{\temp,\kappa}\left(\ve{\theta}_{\kappa},\dve{m}_{\kappa},\ve{z}_{\kappa},\ve{u}_{\kappa},\ve{d}_{\kappa}\right) &= 0,\quad \forall \kappa \in \mathcal{K}, \label{eq:mpc:alg_temp} \\
    \ve{g}_{\temp,\kappa}\left(\ve{\theta}_{\kappa},\dve{m}_{\kappa},\ve{z}_{\kappa},\ve{u}_{\kappa},\ve{d}_{\kappa}\right) &\leq 0,\quad \forall \kappa \in \mathcal{K}. \label{eq:mpc:ineq_temp}
\end{align}
\end{subequations}

Note that splitting the objective into a hydraulic objective $j_{\hydraulic}$ and a temperature objective $j_{\temp}$ does not impose any restrictions, since all optimization variables can still be part of $j_{\temp}$.
Therefore, the optimization problem that must be solved at each time-step $k$ is given by
\begin{align}
\begin{split}
    J^{\star}_k = \min_{\hve{\theta}_k, \hat{\dve{m}}_k, \hve{z}_k, \hve{u}_k} J_k \left(\hve{\theta}_k, \hat{\dve{m}}_k, \hve{z}_k, \hve{u}_k\right) \\ \text{subject to} \ \eqref{eq:mpc:constraintsHydraulic} \text{ and } \eqref{eq:mpc:constraintsTemp}, \label{eq:mpc}
\end{split}
\end{align}
where $J^{\star}_k$ is the optimal function value of \eqref{eq:mpc:objective}.

\subsection{Primal decomposition} \label{sub:primalMPC}
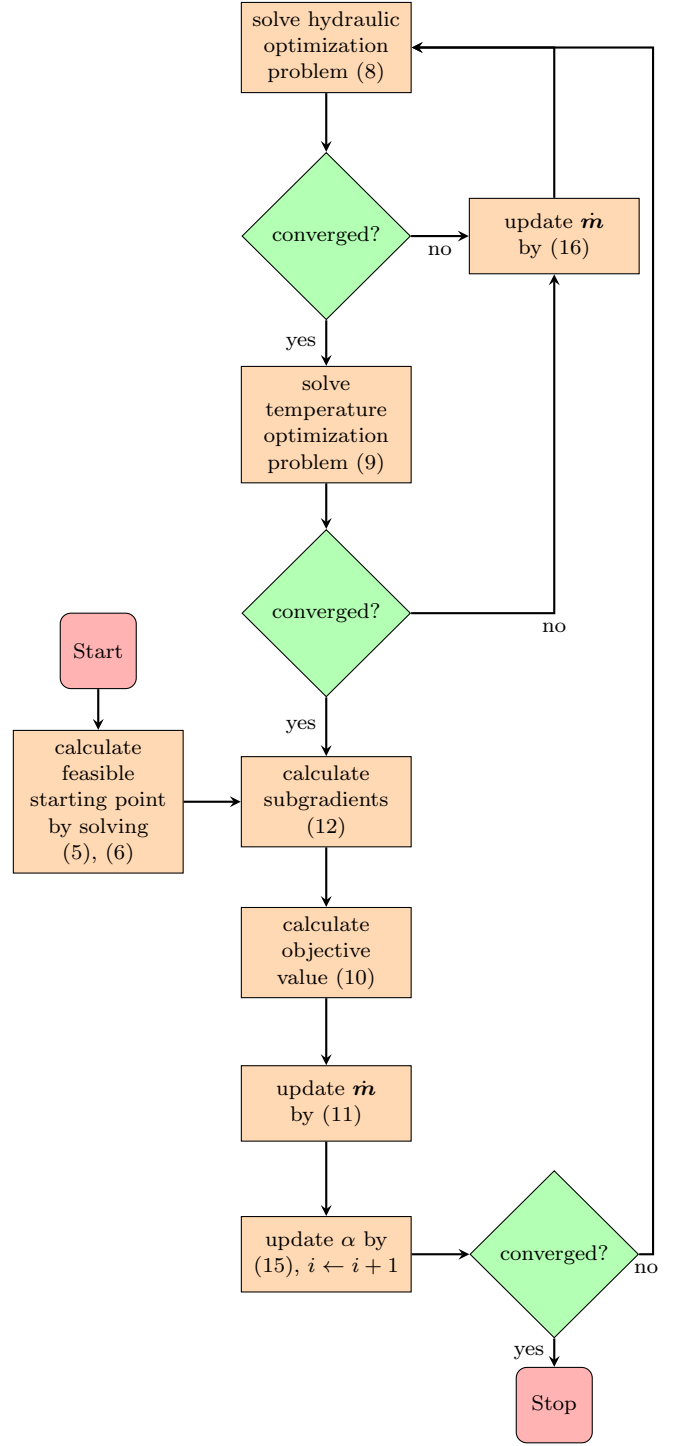
\begin{figure}
    \centering
    \begin{tikzpicture}[node distance=2cm, font=\small]
\node (hydOpt) [process] {solve hydraulic optimization problem \eqref{eq:mpc:hydraulic}};
\node (hc) [decision, below of=hydOpt, yshift=0.5cm] {converged?};
\node (updateFail) [process, right of=hc, xshift=1cm, text width=2cm] {update $\dve{m}$ by \eqref{eq:mflow_REupdate}};
\node (tempOpt) [process, below of=hc, yshift=0.35cm] {solve temperature optimization problem \eqref{eq:mpc:temp}};
\node (tc) [decision, below of=tempOpt, yshift=0.5cm] {converged?};
\node (subgradient) [process, below of=tc, yshift=0.6cm] {calculate subgradients \eqref{eq:subgradient}};
\node (feasStartPoint) [process, left of=subgradient, xshift=-1cm, text width=2cm] {calculate feasible starting point by solving \eqref{eq:mpc:constraintsHydraulic}, \eqref{eq:mpc:constraintsTemp}};
\node (start) [startstop, above of=feasStartPoint, yshift=-0cm] {Start};
\node (fval) [process, below of=subgradient, yshift=0.75cm] {calculate objective value \eqref{eq:primal:objective}};
\node (updateMdot) [process, below of=fval, yshift=1cm] {update $\dve{m}$ by \eqref{eq:mflow_update}};
\node (updateAlpha) [process, below of=updateMdot, yshift=1cm] {update $\alpha$ by \eqref{eq:updateAlpha}, $i\leftarrow i+1$};
\node (converged) [decision, right of=updateAlpha, xshift=1cm] {converged?};
\node (stop) [startstop, below of=converged, yshift=0.5cm] {Stop};
\draw [arrow] (hydOpt) -- (hc);
\draw [arrow] (hc) -- node[anchor=east] {yes} (tempOpt);
\draw [arrow] (hc) -- node[anchor=north] {no} (updateFail);
\draw [arrow] (tempOpt) -- (tc);
\draw [arrow] (tc) -- node[anchor=east] {yes} (subgradient);
\draw [arrow] (tc) -| node[anchor=north] {no} (updateFail);
\draw [arrow] (updateFail) |- (hydOpt);
\draw [arrow] (start) -- (feasStartPoint);
\draw [arrow] (feasStartPoint) -- (subgradient);
\draw [arrow] (subgradient) -- (fval);
\draw [arrow] (fval) -- (updateMdot);
\draw [arrow] (updateMdot) -- (updateAlpha);
\draw [arrow] (updateAlpha) -- (converged);
\draw [arrow] (converged) -- node[anchor=east] {yes} (stop);
\draw [arrow] (converged) -- node[anchor=north] {no} ++ (1.3cm,0) |- (hydOpt); 
\end{tikzpicture}
    \caption{Procedure of the primal decomposition scheme}
    \label{fig:primalProcedure}
\end{figure}

As can be seen in \eqref{eq:timeDiscrete_DE}, assuming $\dve{m}_{k}$ is constant makes \eqref{eq:timeDiscrete_DE} a linear system of equations. If the same holds true for \eqref{eq:timeContinuousAE}--\eqref{eq:timeContinuous_inequality}, then \textit{primal decomposition} can be used to efficiently solve the optimization problem \eqref{eq:mpc}. Figure~\ref{fig:primalProcedure} displays a procedure of the primal decomposition scheme.

First, a feasible starting point is calculated by solving the system of nonlinear equations \eqref{eq:mpc:constraintsHydraulic} and \eqref{eq:mpc:constraintsTemp}, where the control variables are taken from the MPC results of the previous time-step. The algorithm works by iteratively altering $\hat{\dve{m}}_{k}$. The first step in each iteration is to solve the hydraulic optimization problem 
\begin{align}
\begin{split}
    J^{i}_{\hydraulic,k} = \min_{\hve{z}_{\hydraulic,k}, \hve{u}_{\hydraulic,k}} &J_{\hydraulic,k}\left(\hat{\dve{m}}_k^{i}, \hve{z}_{\hydraulic,k}^{i}, \hve{u}_{\hydraulic,k}^{i}\right) \\
    &\text{subject to} \ \eqref{eq:mpc:constraintsHydraulic} \label{eq:mpc:hydraulic}
\end{split}
\end{align}
with $i$ being the primal decomposition iteration index.
Solving \eqref{eq:mpc:hydraulic} yields the optimal values for $\hve{z}_{\hydraulic,k}^{i}$ and $\hve{u}_{\hydraulic,k}^{i}$.

Next, the thermal optimization problem 
\begin{align}
\begin{split}    
    J^{i}_{\temp,k} = \min_{\ve{\theta}_k, \hve{z}_{\temp,k}, \hve{u}_{\temp,k}} &J_{\temp,k}\left(\hve{\theta}_k^{i}, \hat{\dve{m}}_k^{i}, \hve{z}_{\hydraulic,k}^{i}, \hve{z}_{\temp,k}^{i}, \hve{u}_{\hydraulic,k}^{i}, \hve{u}_{\temp,k}^{i}\right) \\
    &\text{subject to} \ \eqref{eq:mpc:constraintsTemp} \label{eq:mpc:temp}
\end{split}
\end{align}
is solved, resulting in $\ve{\theta}_k^{i}$, $\ve{z}_{\temp,k}^{i}$, and $\ve{u}_{\temp,k}^{i}$. The objective value after iteration $i$ is given by
\begin{align}
    J^{i}_{k} = J^{i}_{\hydraulic,k} + J^{i}_{\temp,k}. \label{eq:primal:objective}
\end{align}
A convergence criterion, \eg\ based on $\Delta J=J^{i}_{k} - J^{i-1}_{k}$, can then be used.
If the convergence criterion is not met, the primal variable is updated by \cite{boyd_notes_2008}
\begin{align}
    \hat{\dve{m}}_k^{i+1} = \hat{\dve{m}}_k^{i} - \alpha^i  \ve{s}_k^i \label{eq:mflow_update}
\end{align}
with the subgradient
\begin{align}
    \ve{s}_k = \frac{\partial L_k\left(\hve{\theta}_k, \hat{\dve{m}}_k, \hve{z}_k, \hve{u}_k, \hve{\lambda}_k \right)}{\partial \hat{\dve{m}}_k}, \label{eq:subgradient}
\end{align}
the Lagrangian as in  \cite{palomar_tutorial_2006}
\begin{align}
    L_k = J_k - \sum_{\kappa\in\mathcal{K}} \sum_{\ve{h}\in\mathcal{H}} \transve{\lambda}_{h,\kappa} \ve{h}_{\kappa},
\end{align}
the set of constraint functions $\mathcal{H}=\{\ve{f}_{\hydraulic}, \ve{g}_{\hydraulic}, \ve{f}_{\differential}, \ve{f}_{\temp},\ve{g}_{\temp}\}$,
and the optimal Lagrange multipliers
\begin{align}
    \hve{\lambda}^{\top}_k = \begin{bmatrix}
        \hve{\lambda}_{f_{\hydraulic},k}^{\top} & \hve{\lambda}_{g_{\hydraulic},k}^{\top} & \hve{\lambda}_{f_{\differential},k}^{\top} & \hve{\lambda}_{f_{\temp},k}^{\top} & \hve{\lambda}_{g_{\temp},k}^{\top}
    \end{bmatrix}.
\end{align}
Once again the notation convention from \eqref{eq:mpc:objective_all} is used, \eg\ $\hve{\lambda}_{f_{\hydraulic},k}^{\top}=\begin{bmatrix}
    \ve{\lambda}_{f_{\hydraulic},k}^{\top} & \dots & \ve{\lambda}_{f_{\hydraulic},k+n_{\control}-1}^{\top}
\end{bmatrix}$.
At the end of each iteration, $\alpha^i$ is updated using the nonsummable, diminishing step-size rule \cite{boyd_subgradient_2014}
\begin{align}
    \alpha^{i} = \frac{\alpha}{\sqrt{i}} \label{eq:updateAlpha}
\end{align}
which is in accordance with the conditions specified in \cite{bertsekas_gradient_1997}.

If \eqref{eq:mpc:constraintsHydraulic} or \eqref{eq:mpc:constraintsTemp} become infeasible due to the primal variables update, then the primal variables are re-updated by
\begin{align}
    \hat{\dve{m}}_k^{i} = \hat{\dve{m}}_k^{i-1} - b\alpha^{i-1}  \ve{s}^{i-1}, \label{eq:mflow_REupdate}
\end{align}
and procedure \eqref{eq:mflow_REupdate} is repeated with a decreasing $b<1$ until \eqref{eq:mpc:constraintsHydraulic} and \eqref{eq:mpc:constraintsTemp} are fulfilled. This is similar to the backtracking approach presented by \cite{krishnamoorthy_primal_2019}. \cite{engelmann_scalable_2025} present a different approach, using relaxation techniques and auxiliary variables to find a feasible $\alpha^i$.


\section{UNDERGROUND HEATING SYSTEM}  \label{sec:CaseStudy}

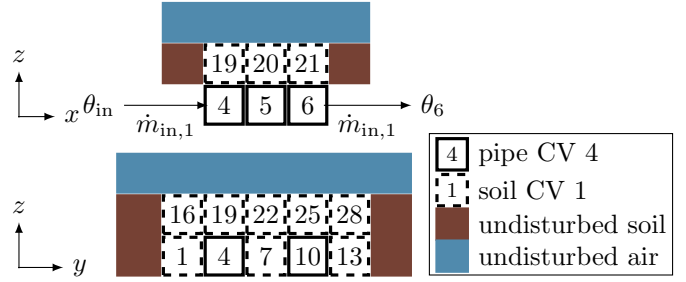
\begin{figure}[t!]
  \centering
  \begin{tikzpicture}
    \draw[black, very thick, dashed] (0,0) rectangle (.5,.5);
    \draw[black, very thick] (.55,0) rectangle (1.05,.5);
    \draw[black, very thick, dashed] (1.1,0) rectangle (1.6,.5);
    \draw[black, very thick] (1.65,0) rectangle (2.15,.5);
    \draw[black, very thick, dashed] (2.2,0) rectangle (2.7,.5);

    \draw[black, very thick, dashed] (0,.55) rectangle (.5,1.05);
    \draw[black, very thick, dashed] (.55,.55) rectangle (1.05,1.05);
    \draw[black, very thick, dashed] (1.1,.55) rectangle (1.6,1.05);
    \draw[black, very thick, dashed] (1.65,.55) rectangle (2.15,1.05);
    \draw[black, very thick, dashed] (2.2,.55) rectangle (2.7,1.05);

    \draw[soiltemp, very thick, fill=soiltemp] (-.6,0) rectangle (-0.05,1.05);
    \draw[soiltemp, very thick, fill=soiltemp] (2.75,0) rectangle (3.25,1.05);
    \draw[airtemp, very thick, fill=airtemp] (-.6,1.1) rectangle (3.25,1.6);
    
    \node (cv1) at (.25,.25) {1};
    \node (cv4) at (.8,.25) {4};
    \node (cv7) at (1.35,.25) {7};
    \node (cv10) at (1.9,.25) {10};
    \node (cv13) at (2.45,.25) {13};
    \node (cv16) at (.25,.8) {16};
    \node (cv19) at (.8,.8) {19};
    \node (cv22) at (1.35,.8) {22};
    \node (cv25) at (1.9,.8) {25};
    \node (cv28) at (2.45,.8) {28};
    \draw[black] (3.5,0) rectangle (6.7,1.875);
    \node[rectangle, draw=black,very thick, minimum size = 4mm] (pipeCV) at (3.8,1.6) {\small 4};
    \node (pipeCVtext) [right=of pipeCV,xshift=-1cm] {pipe CV 4};
    \node[rectangle, draw=black,very thick, minimum size = 4mm,dashed] (soilCV) at (3.8,1.125) {\small 1};
    \node (soilCVtext) [right=of soilCV,xshift=-1cm] {soil CV 1};
    \node[rectangle, draw=soiltemp, fill=soiltemp, very thick, minimum size = 4mm] (soilTemp) at (3.8,.675) {};
    \node (soilTemptext) [right=of soilTemp,xshift=-1cm] {undisturbed soil};
    \node[rectangle, draw=airtemp, fill=airtemp, very thick, minimum size = 4mm] (airTemp) at (3.8,.25) {};
    \node (airTemptext) [right=of airTemp,xshift=-1cm] {undisturbed air};

    \node (axisOrigin) at (-1.9,.01) {};
    \node (axisY) at (-1.1,.1) {$y$};
    \draw[->,>=latex] (-1.912,.1) -- (axisY);
    \node (axisZ) at (-1.9,.9) {$z$};
    \draw[->,>=latex] (-1.9,.089) -- (axisZ);


    \draw[black, very thick] (.55,2) rectangle (1.05,2.5);
    \draw[black, very thick] (1.1,2) rectangle (1.6,2.5);
    \draw[black, very thick] (1.65,2) rectangle (2.15,2.5);

    \draw[black, very thick, dashed] (.55,2.55) rectangle (1.05,3.05);
    \draw[black, very thick, dashed] (1.1,2.55) rectangle (1.6,3.05);
    \draw[black, very thick, dashed] (1.65,2.55) rectangle (2.15,3.05);

    \draw[soiltemp, very thick, fill=soiltemp] (0,2.55) rectangle (.5,3.05);
    \draw[soiltemp, very thick, fill=soiltemp] (2.2,2.55) rectangle (2.7,3.05);
    \draw[airtemp, very thick, fill=airtemp] (0,3.1) rectangle (2.7,3.6);
    
    \node (cv4) at (.8,2.25) {4};
    \node (cv5) at (1.35,2.25) {5};
    \node (cv6) at (1.9,2.25) {6};
    \node (cv19) at (.8,2.8) {19};
    \node (cv20) at (1.35,2.8) {20};
    \node (cv21) at (1.9,2.8) {21};

    \node (m_dot_in) at (-.85,2.25) {$\theta_{\inp}$}; 
    \draw[->,>=latex] (m_dot_in) -- node[anchor=north] {$\dot{m}_{\inp,1}$} (cv4.west);

    \node (m_dot_out) at (3.55,2.25) {$\theta_6$}; 
    \draw[->,>=latex] (cv6.east) -- node[anchor=north] {$\dot{m}_{\inp,1}$} (m_dot_out);

    \node (axisOrigin) at (-1.9,2.1) {};
    \node (axisX) at (-1.2,2.1) {$x$};
    \draw[->,>=latex] (-1.912,2.1) -- (axisX);
    \node (axisZ) at (-1.9,2.9) {$z$};
    \draw[->,>=latex] (-1.9,2.089) -- (axisZ);
\end{tikzpicture}
  \caption{underground heating system with 30 CVs}
  \label{fig:CVpitch}
\end{figure}

\subsection{System model}

To validate the MPC approach and scalability of primal decomposition, we use three versions of a simple underground heating system, commonly applied in underfloor heating and for snow-free surfaces. The versions differ in system size, defined by the number of pipes $n_{\pipe}$, control volumes $n_{\cvcv}$, and CVs per pipe $n_{\mathrm{x}}$ in the $x$-direction. Figure~\ref{fig:CVpitch} shows the system with $n_{\pipe}=2$, $n_{\cvcv}=30$, and $n_{\mathrm{x}}=3$. All cases assume two CV layers in the $z$-direction and soil CVs at both ends in the $y$-direction. The goal is to minimize energy costs while maintaining soil CV temperatures between \SI{278.15}{\kelvin} and \SI{313.15}{\kelvin}.

\subsubsection{System variables}
The states of the system are the CV temperatures $\transve{\theta}=\begin{bmatrix}
    \theta_1 & \dots & \theta_{n_{\cvcv}}
\end{bmatrix}$. The algebraic states are the CV mass flows $\dve{m}^{\top}=\begin{bmatrix}
    \dot{m}_1 & \dots & \dot{m}_{n_{\mathrm{x}}}
\end{bmatrix}$, the pipe pressure losses $\transve{\Delta p}=\begin{bmatrix}
    \Delta p_1 & \dots & \Delta p_{n_{\pipe}}
\end{bmatrix}$, the input temperature of the pipes $\theta_{\inp}$ and the temperature $\theta_{\out}$ of the mass flow after mixing all pipe exit temperatures (see \eqref{eq:tempMixing}). The controllable inputs of the system consist of the mass flows in the pipes $\dve{m}_{\inp}^{\top}=\begin{bmatrix}
    \dot{m}_{\inp,1} & \dots & \dot{m}_{\inp,n_{\cvcv}}
\end{bmatrix}$ and the temperature increase $\Delta \theta$ in between pipe out- and input. The uncontrollable inputs are soil temperature $\theta_{\soil}$ and air temperature $\theta_{\air}$.
Table~\ref{tab:vars} depicts all variables together with their assignment to the vectors contained in $\mathcal{V}$.

\subsubsection{Objective}
The goal of the controller is to minimize the cost of the required electrical energy for the pumps and the heating system given the electricity price $c_{\elec,k}$. The cost of electricity for the pumps per time step is given by
\begin{subequations} \label{eq:pitch:objective}
\begin{align}
    j_{\hydraulic,k} = \frac{c_{\elec,k}}{\rho_{\water} \eta_{\mathrm{pu}}} \transve{z}_{\hydraulic,k} \ve{u}_{\hydraulic,k}
\end{align}
with the pump efficiency $\eta_{\mathrm{pu}}=\num{0.8}$. Assuming an electric boiler is used, the cost of the electricity required to heat the water between the pipe inputs and outputs is given by
\begin{align}
     j_{\temp, k} &= c_{\elec,k} c_{\water} u_{\temp,k} \ma{1}_{1\times n_{\pipe}} \ve{u}_{\hydraulic,k}.
\end{align}
\end{subequations}

\subsubsection{System dynamics}
The system dynamics are based on \eqref{eq:systemDynSimple} and are not repeated here due to space limitations. The parameters are set to $V=\SI{3.75}{\cubic \meter}$, $\sigma_{\mathrm{x}}=\SI{0.025}{\watt \per\kelvin}$, $\sigma_{\mathrm{y}}=\sigma_{\mathrm{z}}=\SI{22.5}{\watt \per\kelvin}$, $\sigma_{\pipe}=\SI{44.86}{\watt \per \kelvin}$, $c_{\water}=\SI{4200}{\joule \per \kilo \gram \per \kelvin}$, $\rho_{\water}=\SI{1000}{\kilo \gram \per \cubic \meter}$, and $c_{\soil}\rho_{\soil}=\SI{1.5e6}{\joule \per \cubic \meter \kelvin}$.

\newcolumntype{K}[1]{>{\centering\arraybackslash}p{#1}}

\begin{table}[t!]
\begin{center}
\caption{System variables}\label{tab:vars}
\begin{tabular}{K{0.5cm}K{0.5cm}K{0.5cm}K{0.5cm}|K{0.85cm}K{0.85cm}|K{1.7cm}}
\toprule
\multicolumn{4}{c|}{states} & \multicolumn{2}{K{1.7cm}|}{controllable inputs} & uncontrollable inputs  \\
\midrule
$\ve{\theta}$ & $\ve{z}_{\hydraulic}$ & $\ve{z}_{\temp}$ & $\dve{m}$ & $\ve{u}_{\hydraulic}$ & $\ve{u}_{\temp}$ & $\ve{d}$ \\
\midrule
$\theta_1$ & $\Delta p_1$ & $\theta_{\inp}$ & $\dot{m}_1$ & $\dot{m}_{\inp,1}$ &  & $\theta_{\soil}$ \\
$\vdots$ & $\vdots$ & & $\vdots$ & $\vdots$ & $\Delta \theta$ &  \\
$\theta_{n_{\cvcv}}$ & $\Delta p_{n_{\pipe}}$ & $\theta_{\out}$ & $\dot{m}_{n_{\mathrm{x}}}$ & $\dot{m}_{\inp,n_{\pipe}}$ &  & $\theta_{\air}$ \\
\bottomrule
\end{tabular}
\end{center}
\end{table}

\subsubsection{Algebraic equations}
In series with each pipe $p$, there is a pump that controls the mass flow rate $\dot{m}_{\inp,p}$.
The first set of algebraic equations are the pressure loss relations 
\begin{subequations} \label{eq:pitch:algebraic}
    \begin{align}
        \Delta p_p\left( \dot{m}_{\inp,p} \right) = \frac{\Delta p_{\nominal}}{\dot{m}_{\nominal}^2} \dot{m}_{\inp,p}^2,
    \end{align}
\end{subequations}
where $\Delta p_{\nominal}=\SI{8e5}{\pascal}$ and $\dot{m}_{\nominal}=\SI{30}{\kg \per \second}$ are nominal values.  The input temperature is given by
\begin{align}
   \theta_{\inp} = \theta_{\out} + \Delta \theta \label{eq:tempDifference}
\end{align}
and $\theta_{\out}$ is governed by
\begin{align}
   \left(\sum_{p=1}^{n_{\pipe}} \dot{m}_{\inp,p} \right) \theta_{\out} = \left(\sum_{p=1}^{n_{\pipe}} \dot{m}_{\inp,p} \theta_{\out,p} \right), \label{eq:tempMixing}
\end{align}
where $\theta_{\out,p}$ is the temperature of the last CV of pipe $p$, for example $\theta_6$ for $p=1$ and $\theta_{12}$ for $p=2$ in Figure~\ref{fig:CVpitch}.

\subsection{Time discretization}
\begin{figure}
\centering
\begin{subfigure}{\columnwidth}
    \input{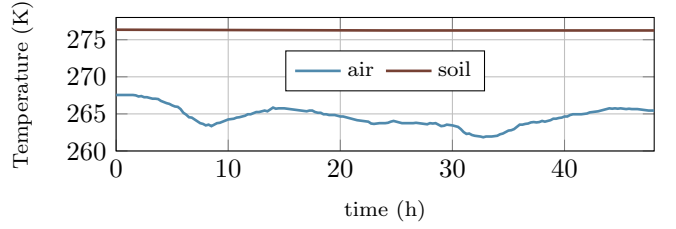}
    \caption{ambient temperatures}
    \label{fig:first}
\end{subfigure}
\hfill
\begin{subfigure}{\columnwidth}
    \input{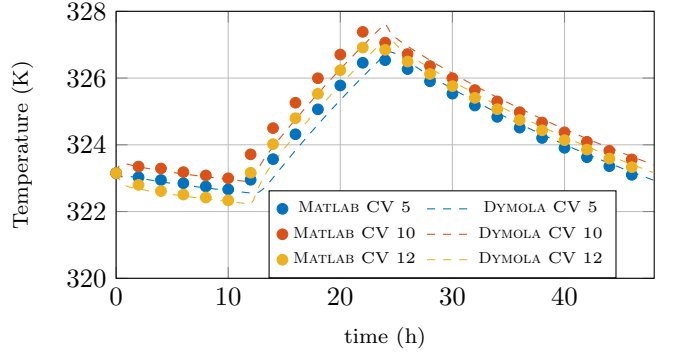}
    \caption{pipe temperatures}
    \label{fig:second}
\end{subfigure}
\hfill
\begin{subfigure}{\columnwidth}
    \input{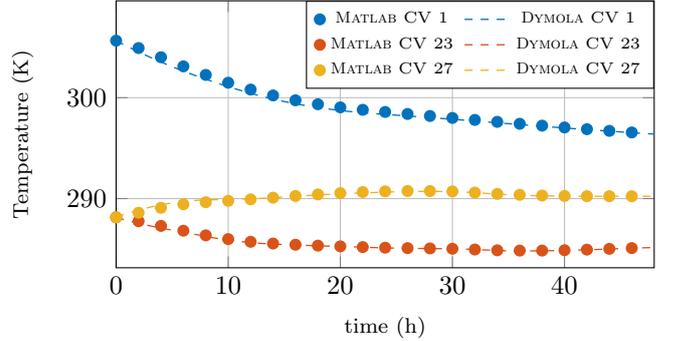}
    \caption{soil temperatures}
    \label{fig:third}
\end{subfigure}
        
\caption{Comparison of the \textsc{Dymola} (dashed lines) and the \textsc{Matlab} (dots) simulations for the system in Figure~\ref{fig:CVpitch}}
\label{fig:dymolaDis}
\end{figure}

One version of the system was implemented in \textsc{Matlab} using the time discretization from \eqref{eq:timeDiscrete_DE} with a fixed time step of $\Delta t =\SI{2}{\hour}$, 
and simulated by solving the resulting equations. A second version, created in \textsc{Dymola} with the library from \cite{westphal_enabling_2025}, also uses backward differentiation but adaptively selects $\Delta t$ and differentiation order based on the system state.
Due to the validated models and advanced discretization, the \textsc{Dymola} model serves as the benchmark.
Figure~\ref{fig:dymolaDis} compares both simulations, showing the \textsc{Matlab} model is more accurate for soil CVs (maximum error \SI{0.473}{\kelvin}) than pipe CVs (maximum error \SI{0.936}{\kelvin}) due to slower soil dynamics.
Largest pipe deviations occur after increasing $\dot{m}_{\inp,2}$ from \SI{2}{\kilo \gram \per \second} to \SI{5}{\kilo \gram \per \second} at \SI{10}{\hour} and decreasing $\Delta \theta$ from \SI{1}{\kelvin} to \SI{0.5}{\kelvin}. Figure~\ref{fig:CVpitch} shows ambient temperature data; $\dot{m}_{\inp,1}$ is held constant at \SI{2}{\kilo \gram \per \second}. For higher accuracy, $\Delta t$ must be reduced; at $\Delta t=\SI{15}{\minute}$, the maximum pipe temperature error decreases to \SI{0.23}{\kelvin}.

\begin{table}
\begin{center}
\caption{Numerical results of the two MPCs}\label{tab:res}
\begin{tabularx}{\columnwidth}{XXXXXX}
\toprule
number of pipes & number of variables & \multicolumn{2}{>{\hsize=\dimexpr2\hsize+2\tabcolsep+\arrayrulewidth\relax}X}{average $j_{\hydraulic}$ + $j_{\temp}$ per time step in \unit{\sieuro}} & \multicolumn{2}{>{\hsize=\dimexpr2\hsize+2\tabcolsep+\arrayrulewidth\relax}X}{average calc. time per time step in \unit{\second}} \\
& & IPM MPC & PD MPC & IPM MPC & PD MPC \\
\midrule
1 & 19 & \num{116.01} & \num{117.91} & \num{1.18} & \num{6.03} \\
2 & 43 & \num{271.53} & \num{274.4} & \num{12.51} & \num{7.81} \\ 
7 & 202 & \num{1074.7} & \num{1298.5} & \num{3706} & \num{124.29} \\ 
\bottomrule
\end{tabularx}
\end{center}
\end{table}

\subsection{Model predictive controllers}
Two MPCs were implemented in \textsc{Matlab} with a control-horizon of $n_{\control}=12$. One solved optimization problem \eqref{eq:mpc} using \textsc{Matlab}’s standard interior-point method (IPM), while the other used the primal decomposition (PD) scheme introduced here with primal variables $\ve{u}_{\hydraulic}$.


\subsection{Results of the model predictive controller}
As shown in Table~\ref{tab:res}, the IPM outperformed the PD in all three system setups in terms of objective value. However, the PD outperformed the IPM in terms of calculation time for all but the smallest system setup. For $n_{\pipe}=2$, the PD MPC calculation time is \SI{62.4}{\percent} of the IPM MPC calculation time; for $n_{\pipe}=7$, it is \SI{3.35}{\percent}. In conclusion, the PD MPC is more scalable than the IPM MPC. For the two smallest system setups, the average cost per time step is nearly equal for both MPCs, and for the system with seven pipes, the IPM MPC outperforms the PD MPC by \SI{17.2}{\percent}.

\begin{figure}[t!]
    \centering
    \definecolor{mycolor1}{rgb}{0.00000,0.44700,0.74100}%
\definecolor{mycolor2}{rgb}{0.85000,0.32500,0.09800}%
\definecolor{mycolor3}{rgb}{0.92900,0.69400,0.12500}%

\begin{tikzpicture}
        \def\hpmark{pentagon}
        \def\dhnmark{diamond}
        \def\gasmark{asterisk}
        \def\othermark{x}
        \begin{groupplot}[
            group style={
                group size=1 by 5,
                x descriptions at=edge bottom,
                vertical sep=0pt,
            },
            xmode=normal,
            xlabel= {time (\unit{\hour})},
            xmajorgrids=true,
            xminorgrids=true,
            ymajorgrids=true,
            width=\columnwidth,
            height=0.16\textheight,
            ytick pos=left,
            xmin=0,
            xmax=48,
            label style={font=\small},
        ]
        \pgfplotsset{scaled y ticks=false}
        \nextgroupplot[
            ymode=normal,
            legend style={font=\tiny, at={(0,1)},anchor=north west},
            ylabel={temperature (\unit{\kelvin})},
        ]
            \input{mpcPipeTemp}
    
        \nextgroupplot[
            ymode=normal,
            ylabel= {temperature (\unit{\kelvin})},
            legend columns=4,
            legend style={/tikz/column 4/.style={
                            column sep=2pt,
                        },font=\tiny, at={(1,1)},anchor=north east},
            ymin=275,
            ymax=313,
        ]
            \input{mpcsoilTemp}
        \nextgroupplot[
            height=0.14\textheight,
            ymode=normal,
            ylabel={mass flow (\unit[per-mode=fraction]{\kilo \gram \per \second})},
            legend style={font=\tiny, at={(0,1)},anchor=north west},
        ]
            \input{mpcMflow}
        \nextgroupplot[
            height=0.12\textheight,
            ymode=normal,
            ylabel={$\Delta \theta$ (\unit{\kelvin})},
        ]
            \input{mpcDeltaTheta}
        \nextgroupplot[
            height=0.14\textheight,
            ymode=normal,
            ylabel = {elec. price (\unit[per-mode=fraction]{\sieuro \per \kilo \watt \per\hour})},
            ymin=0,
            ymax=33,
        ]
            \input{elecPrice}
        \end{groupplot}
    \end{tikzpicture}
    \caption{Primal MPC results for the system in Figure~\ref{fig:CVpitch}}
    \label{fig:mpc}
\end{figure}

Figure~\ref{fig:mpc} shows the simulation results for the system controlled by the PD MPC. When electricity prices are low, the MPC increases $\Delta \theta$, resulting in higher pipe temperatures. Otherwise, using only the mass flow rate is sufficient for distributing the stored heat at the beginning of the simulation without raising the temperature. As the soil temperature decreases, the mass flow increases. Note that the same mass flow is used for both pipes due to the symmetry of the system. The simulation results clearly demonstrate that the PD MPC can control the system according to changing electricity prices and leverage the system's flexibility to minimize operating costs.

\subsection{Discussion of the MPC results}

Assuming constant fluid properties is common for similar systems (\eg\ \cite{maurer_toward_2023}). However, incorporating temperature-dependent fluid properties into the primal decomposition would enhance system accuracy. To fully evaluate the benefits of primal decomposition, more advanced system setups that account for uncertainty in predictions are required. Additionally, a detailed analysis of control quality and convergence guarantees concerning the time step $\Delta t$ and control horizon is necessary.
\section{CONCLUSION}  \label{sec:Conclusion}
This work presents a direct method for model predictive controller design for thermo-hydraulic systems modeled by control volumes. Using primal-decomposition increases the approach's scalability massively compared to a standard MPC approach. Future work should include a more extensive analysis of more sophisticated system setups to provide a more robust evaluation of the presented approach. Additionally, comparisons with other automated MPC frameworks, as presented in \cite{houska_acado_2011} and \cite{chen_matmpc_2019}, are necessary.

\section*{DECLARATION OF GENERATIVE AI AND AI-ASSISTED TECHNOLOGIES IN THE WRITING PROCESS}
During the preparation of this work the authors used DeepL Write and GPT4.1-Mini in order to improve readability. After using these tools, the authors reviewed and edited the content as needed and take full responsibility for the content of the publication.

\bibliography{TurfMPC}            
                                                   







\appendix
\end{document}